\documentclass[pasp]{emulateapj}

\usepackage{natbib}
\usepackage{multirow}
\usepackage{appendix}
\usepackage{longtable}

\slugcomment{PASP accepted}

\shorttitle{NIR Imaging Polarimetry with NICMOS}
\shortauthors{Batcheldor et al.}

\begin{document}


\title{High Accuracy Near-infrared Imaging Polarimetry with NICMOS\altaffilmark{1}}

\author{D. Batcheldor,\altaffilmark{2} G. Schneider,\altaffilmark{3} D. C. Hines,\altaffilmark{4} G. D. Schmidt,\altaffilmark{3} 
D. J. Axon,\altaffilmark{5}\\ A. Robinson,\altaffilmark{5} W. Sparks\altaffilmark{6} \& C. Tadhunter\altaffilmark{7}}
\email{dan@astro.rit.edu}










\altaffiltext{1}{Based on observations made with the NASA/ESA Hubble Space Telescope obtained at the 
Space Telescope Science Institute, which is operated by the Association of Universities for Research in Astronomy, Incorporated, under 
NASA contract NAS 5-26555.}

\altaffiltext{2}{Center for Imaging Science, Rochester Institute of Technology, 54 Lomb Memorial Drive, Rochester, NY, 14623, USA}
\altaffiltext{3}{Steward Observatory, The University of Arizona, 933 N. Cherry Avenue, Tucson, AZ, 85721, USA}
\altaffiltext{4}{Space Science Institute, 4750 Walnut Street, Suite 205, Boulder, CO 80301, USA}
\altaffiltext{5}{Department of Physics, Rochester Institute of Technology, 54 Lomb Memorial Drive, Rochester, NY, 14623, USA}
\altaffiltext{6}{Space Telescope Science Institute, 3700 San Martin Drive, Baltimore, MD, 21218, USA}
\altaffiltext{7}{Department of Physics \& Astronomy, University of Sheffield, Sheffield, S3 7RH, UK}

\begin{abstract}
The findings of a nine orbit calibration plan carried out during {\it HST} Cycle 15, to fully determine the NICMOS 
camera 2 (2.0\micron) polarization calibration to high accuracy, are reported. Recently Ueta et al. and Batcheldor 
et al. have suggested that NICMOS possesses a residual instrumental polarization at a level of $1.2-1.5\%$. This 
would completely inhibit the data reduction in a number of GO programs, and hamper the ability of the instrument 
to perform high accuracy polarimetry. We obtained polarimetric calibration observations of three polarimetric 
standards at three spacecraft roll angles separated by $\sim60\degr$. Combined with archival data, these 
observations were used to characterize the residual instrumental polarization in order for NICMOS to reach its 
full potential of accurate imaging polarimetry at $p\approx1\%$. Using these data, we place an 0.6\% upper limit 
on the instrumental polarization and calculate values of the parallel transmission coefficients that reproduce 
the ground-based results for the polarimetric standards. The uncertainties associated with the parallel 
transmission coefficients, a result of the photometric repeatability of the observations, are seen to dominate 
the accuracy of $p$ and $\theta$. However, the updated coefficients do allow imaging polarimetry of targets with 
$p\approx1.0\%$ at an accuracy of $\pm0.6\%$ and $\pm15\degr$. This work enables a new caliber of science with 
{\it HST}.
\end{abstract}


\section{Introduction}

Polarimetry is a powerful observational tool that augments and complements the capabilities of imaging, photometry and spectroscopy. 
While the latter allow the determination of spatial distributions, chemical composition and dynamics, polarimetry lets us observe 
the nature of magnetic fields, object orientation, scattering, and the properties of interstellar particles in general. It also allows 
us to probe the nature of emission mechanisms (e.g., synchrotron/thermal), and to investigate the geometry of unresolved sources. 

The value of polarimetry to elucidate the physical nature of light-scattering particles and their environments in astrophysical systems 
has been repeatedly demonstrated in, for example, active galactic nuclei \citep{1985ApJ...297..621A,1995ApJ...452L..87C}, galactic and 
extra-galactic magnetic fields \citep{1995RvMA....8..185W}, radio galaxies \citep{1998A&A...331..475F,2001A&A...366....7V}, and circumstellar 
disks \citep{2001ApJ...553L.189K}. However, because of ``beam depolarization'', results derived from polarimetric imaging observations may 
be biased when obtained at low spatial resolution. For example, compact (but spatially resolved) objects that exhibit circular symmetry or 
bi-conical structures, like active galactic nuclei (AGN) and stars surrounded by orbiting circumstellar dust, contain polarized elements 
that partially reduce, if not fully cancel, each other when spatially averaged. If the object is unresolved, then the polarization vectors 
sum to {\it zero}; in any case, beam depolarization will always underestimate the intrinsic polarization. 

NICMOS is the {\it only} near infrared (NIR) instrument capable of the high resolution, high fidelity polarimetry needed to examine the scattering geometry 
and materials, in detail, for many types of astronomical objects. To date, there have been a multitude of studies carried out using NICMOS direct imaging 
polarimetry \citep[e.g.][]{2000ApJ...544..269C,2000MNRAS.313L..52T,2000ApJ...536L..89S,2002ApJ...576..429S,2002ApJ...574...95S,2003AJ....126..848S,
2005ApJ...634.1146M,2006ApJ...642..339S} and there are now more ambitious programs to measure the characteristics of AGN and regions within circumstellar 
protoplanetary debris disks, that typically exhibit polarizations of less than 5\% \citep{2004MNRAS.350..140S,2005AAS...207.1013S}. However, there have been 
recent reports of a systematic residual instrumental polarization of $\sim1.5\%$ in some objects \citep{2005AJ....129.1625U}, which although acceptable for 
highly polarized targets, seriously compromises studies of low polarization objects. A comprehensive study of this residual polarization has been carried out 
by \citet[][hereafter B06]{2006PASP..118..642B}. Using all of the available standard polarization data in the archives, B06 indeed find a residual excess polarization 
of $\approx1.2\%$ in the NIC2 camera. The extent of this excess remains unknown for NIC1. 

Unfortunately, prior to the new calibration reported in this paper, since the advent of the NICMOS cooling system, NCS \citep{2004AdSpR..34..543S}, installed during 
SM3B in 2002, observations of only one polarized and one unpolarized standard have been obtained, and each at only two celestial orientation angles. Hence, it 
has been impossible, with those limiting data, to entirely characterize the NICMOS residual polarization. Polarization measurements at three well separated roll 
angles are required to remove the dependence of the measured Stokes parameters on the relative transmission of each of the filters. 

\begin{deluxetable}{lcccc}
\tablecaption{Details of Observations \label{tab:obs}}
\tablewidth{0pt}
\tablehead{
\colhead{Target}&\colhead{DATE-OBS}&\colhead{ORIENTAT}&\colhead{Camera}&\colhead{T (s)} 
}
\startdata
VR 84c          & 2007-05-03 & 182.2\degr & NIC1 &  3.6 \\
		& 2007-08-03 & 299.3\degr &      &      \\
		& 2006-12-15 &  59.2\degr &      &      \\   
		& 2007-05-03 & 181.4\degr & NIC2 & 14.0 \\
		& 2007-08-03 & 298.6\degr &      &      \\
		& 2006-12-15 &  58.4\degr &      &      \\ 
VSS VIII-13     & 2006-08-04 & 180.3\degr & NIC1 &  1.6 \\
		& 2006-06-25 & 114.3\degr &      &      \\
		& 2007-04-11 &  54.3\degr &      &      \\
                & 2006-08-04 & 179.6\degr & NIC2 &  7.0	\\
		& 2006-06-25 & 113.6\degr &      &      \\
		& 2007-04-11 &  53.6\degr &      &      \\
HD331891        & 2006-08-27 & 273.8\degr & NIC1 &  1.6 \\
		& 2006-07-01 & 338.4\degr &      &      \\
		& 2007-04-15 &  38.3\degr &      &      \\
                & 2006-08-27 & 273.1\degr & NIC2 &  7.0 \\
		& 2006-07-01 & 337.6\degr &      &      \\
		& 2007-04-15 &  37.5\degr &      &      \\
\enddata
\tablecomments{Each observation was performed at three different roll angles and for the POL0, POL120 and 
POL240 polarizers. In total, NIC2 was dithered around 36 pointings per orbit, with 12 pointings 
per polarizer. The DATE-OBS parameter gives the observation date in YYYY-MM-DD format. 
``ORIENTAT'' refers to the celestial position angle of the image +y axis (degrees east of north). 
``T'' gives the exposure times in seconds.}
\end{deluxetable}

This paper reports the findings of a nine orbit {\it HST} calibration plan (Cycle 15), to fully investigate the residual polarization in NICMOS. NICMOS carries two 
cameras with polarimetry optics designated NIC1 and NIC2 (a third camera for wide-field and grism imaging has no polarimetry capability). NIC1 provides an 
11''$\times$11'' field-of-view (FOV) with an image scale of 43.1 milli-arcsecs (mas) per pixel and 1.045\micron~ broadband (0.475\micron~ FWHM; $R=\Delta\lambda/\lambda = 45\%$) 
polarizing filters. NIC2 (the primary focus of this paper) provides a 19\farcs3$\times$19\farcs2 FOV with 75.8 mas pixels and a medium bandwidth polarimetric passband 
of 1.994\micron~ (0.202\micron~ FWHM; $R=10\%$). Both optical channels critically sample the respective point spread functions (PSFs) in their polarimetric passband. For 
additional details of the NICMOS instrument see \cite{1998ApJ...492L..95T} and the NICMOS Instrument Handbook \citep{2007hsti.book.....B}. The two cameras have non common 
path optics from the instrument's field divider mirror assembly to their respective re-imaged focal plane detectors. As polarimetric imaging in the two cameras is (uniquely) 
carried out at different wavelengths, there is no reason to expect the same level of residual instrumental polarization in each camera. With this in mind we carried out observing 
plans to push the NIC2 calibration to its instrumental limit, given the intrinsically superior polarimetric performance of the NIC2 camera over NIC1 \citep{2000PASP..112..983H}. 
Due to the nature of NIC1, we confine the results from this camera to Appendix~\ref{app:a}. In \S~\ref{obs} we describe the observing strategy employed and in \S~\ref{red} we 
explain the data reduction procedures. A detailed examination of the photometry is completed in \S~\ref{phot}. The results and methods are presented in \S~\ref{ptheta} before 
being discussed in \S~\ref{dis} and concluded in \S~\ref{cons}. A recommended observing plan for NICMOS imaging polarimetry with NIC2 is included in \S~\ref{dis2}. 

\section{Observations}\label{obs}

\begin{figure}
\plotone{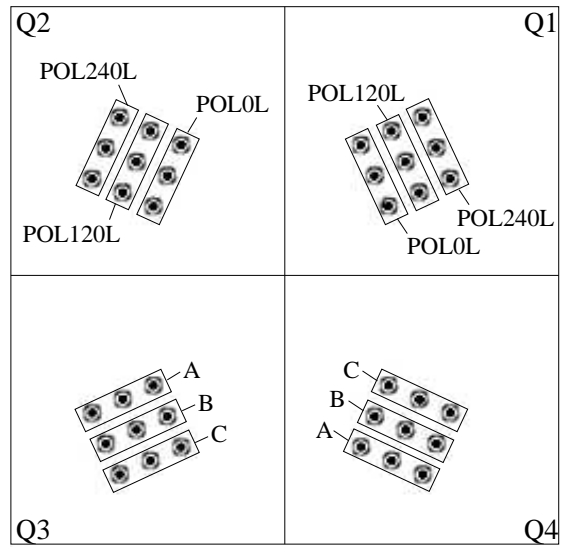}
\caption[NIC2 Pointings]{
An example from one orbit of the NIC2 pointings used to determine the effects of the IPRFs, quadrant dependence and persistence. POL0L, POL120L and POL240L 
observations are in the inner, middle and outer ``diamonds'' of points respectively. The ``A'', ``B'' and ``C'' positions are used to index the dithered pointings 
of each polarizer. The POL0L, POL120L, POL240 and A, B, C markings, that are both applicable to each quadrant, are used to define the pointing naming 
convention. For example, the lowest pointing on the left of the array would be POL240L-Q3-C.
}\label{fig:nic2point}
\end{figure}

The NIC2 observations were designed to be totally comprehensive in the sense that they were not only made to address the NICMOS residual polarization, 
but several other key issues including possible dependence with detector quadrants, source color dependence, the inter-pixel response functions (IPRFs) 
and latent image persistence. The details of the observations are summarized in Table~\ref{tab:obs}. 

Image persistence (or latent image decay) can contribute significantly to photometric errors in sequential exposures of bright targets on the same pixel locations 
on the focal plane array. The same is also true for cosmic ray hits, therefore all observations were scheduled in orbits that were not impacted by the South Atlantic 
Anomaly. As NICMOS polarimetric analysis derives intensity differences from measures obtained in each of its three camera-specific polarizing filters, systematic 
differences due to image persistence, or other instrumental causes, should be circumvented or minimized to the greatest degree possible. Persistence can be overcome 
with an effective image dithering strategy. The diffraction spikes arising from the {\it HST} secondary mirror support in the {\it HST}+NICMOS PSF can contain a 
large number of photons, so a precise dithering pattern has to be carefully chosen. In our calibration observations, custom dither patterns were created for each camera. 

\begin{deluxetable*}{lccccccccc}[t]
\tablecaption{Details of Polarimetric Standards \label{tab:stand}}
\tablecolumns{10}
\tablewidth{0pt}
\tablehead{
\colhead{Target}&\colhead{Type}        &\colhead{$m_v$}&\colhead{$m_j$}&\colhead{$m_k$} 
&\colhead{0.55\micron} &\colhead{1.05\micron} &\colhead{1.21\micron} &\colhead{2.00\micron} &\colhead{2.04\micron}       \\
&                      &               &               &                
&\multicolumn{5}{c}{$(p\%,\theta)$}
}
\startdata
VR 84c (WKK F7) & B5V   & 10.3 & 9.34 & 9.13 & 5.98,118\degr & 4.04,118\degr* & 3.19,118\degr & 1.25,126\degr* & 1.19,126\degr \\   
VSS VIII-13     & K1III & 12.7 & 9.65 & 8.68 & 2.25,102\degr & 2.49,99\degr*  & 2.21,96\degr  & 0.91,102\degr* & 0.86,102\degr \\
HD 331891       & A4III & 9.3  & 8.80 & 8.72 & 0.04,n/a      & \nodata        & \nodata       & \nodata        & \nodata      
\enddata
\tablecomments{
Basic data for the observed polarimetric standards. Two polarized and one un-polarized star (HD331891). Values marked with ``*'' are those that have been 
corrected using the Serkowski curve. There is no method to precisely determine $\theta$ at $1.05\micron$ and $2.00\micron$ so the average values between 
$0.55\micron$ and $1.21\micron$, and the values at $2.04\micron$ are used respectively. More details for each target are given in the text. 
}
\end{deluxetable*}

In the absence of super-dark calibration frames, that replicate to very high fidelity the behavior of the NICMOS detectors as specifically clocked for any given science 
observation, both quadrant dependent and global bias offsets often arise in basic (pipeline) calibrations (both are frequently seen when calibration database/library 
or synthetic dark frames are employed). However, during this calibration we acquired contemporaneous dark frames to mitigate the possibility of residual systematics 
in the multi-accum derived count rate images. These darks were matched to the detector clocking modes used for our targeted observations, and executed during the 
occulted period of each orbit. None-the-less, to assess the possible impact of ``pedestal-like'' effects of other instrumental origins, the same observing sequence 
was repeated with the target repositioned in each of the four detector quadrants. Flat-field artifacts can be enhanced (rather than reduced) in amplification with 
improperly biased (subtracted) dark reference frames.

The observations in each quadrant were dithered in order to mitigate the effects of the IPRFs and to sample around defective pixels. The IPRFs can alter the count rates 
from a target simply through different alignments (centerings) with respects to the pixel center at each pointing. The pointing offsets cover a phase spacing of -1/3, 0 
and +1/3 pixels in order to best tile the IPRFs for the dither pattern employed. In addition, the dither step sizes are greater than the radius of the $3^{\rm rd}$ Airy 
PSF minima (0\farcs58, 7.6 pixels). The orientation of the dither pattern was rotated, with respect to the detector rows and columns, by 22.5\degr~to remove the possibility 
of overlapping persistence from the PSF diffraction spikes of previous pointings. Figure~\ref{fig:nic2point} shows an overlay of the NIC2 pointings from within one orbit. 

These observations were repeated for each of the three chosen polarimetric standard stars and at three separate spacecraft roll angles, i.e., three celestial orientation 
angles of the camera aperture on the sky. This enables a unique determination of any residual instrumental polarization, above the detection floor, relative to the equatorial 
reference frame. More crucially, this multi-orientation technique means that the polarimetric analysis derived from the three polarizers at each of the field orientations can 
be completely and uniquely decoupled and checked for consistency. This method also removes the dependence of the measured Stokes parameters on the relative transmission 
of each of the polarizing filters and, at the same time, any dependence of the instrumental polarization on the detector's sensitivity to polarized light.

Light scattering within the instrument could, at least in part, be largely responsible for the observed residual instrumental polarization, but light 
scattering is wavelength dependent, so standard stars with intrinsically different spectral energy distributions were chosen. The details of the three 
polarimetric standards picked for this calibration are summarized in Table~\ref{tab:stand}. As {\it HST} was operating in two gyro during {\it HST} Cycle 
15 (the epoch of all the calibration observations - see Table~\ref{tab:obs}), the dominant factor in the selection of this sample was the availability 
of the three separate roll angles required to allow polarization measurements to be derived from the single polarizers. 

\subsection{Calibration Targets}

All targets were sufficiently bright so that the PSF central pixel approached the full-well depth near the end of the non-destructive sequence of readouts 
in each multiaccum exposure. This maximized the photon signal to noise ratio with the greatest observing efficiency. In addition, as the exposure times 
were short (Table~\ref{tab:obs}), the target brightnesses ensured each could be observed multiple times within its dither pattern during one orbit. 

VR 84c (WKK F 7, Cha DC F7) was chosen as it has a significant percentage polarization in both the J and K bands, and as it had been observed in the previous 
polarimetric calibrations of NICMOS; it serves as a control. \citet{1992ApJ...386..562W} give a K-band polarization of $1.19\pm0.01\%$ at $126\pm4\degr$. 

The second polarimetric standard VSS VIII-13 (R CrA DC No.13) was also chosen for its percentage polarization. However, it is also found in a different 
association from VR 84c making it more suitable as a calibration target. Within any given association, measured stellar polarization angles tend to be 
similar. This implies a polarization process internal to the association. \citet{1992ApJ...386..562W} give a K-band polarization of $0.9\pm0.2\%$ at $102\pm5\degr$. 

The final standard, HD 331891, was chosen as it was found to have an insignificant amount of polarization at 0.55\micron~($0.04\pm0.02\%$) by \citet{1990AJ.....99.1243T}.
It too was observed in previous calibration projects. There are no data of this target at 2.00\micron (previous to this study), therefore we have to cautiously assume that 
it remains unpolarized at this wavelength. 

\begin{figure*}
\plotone{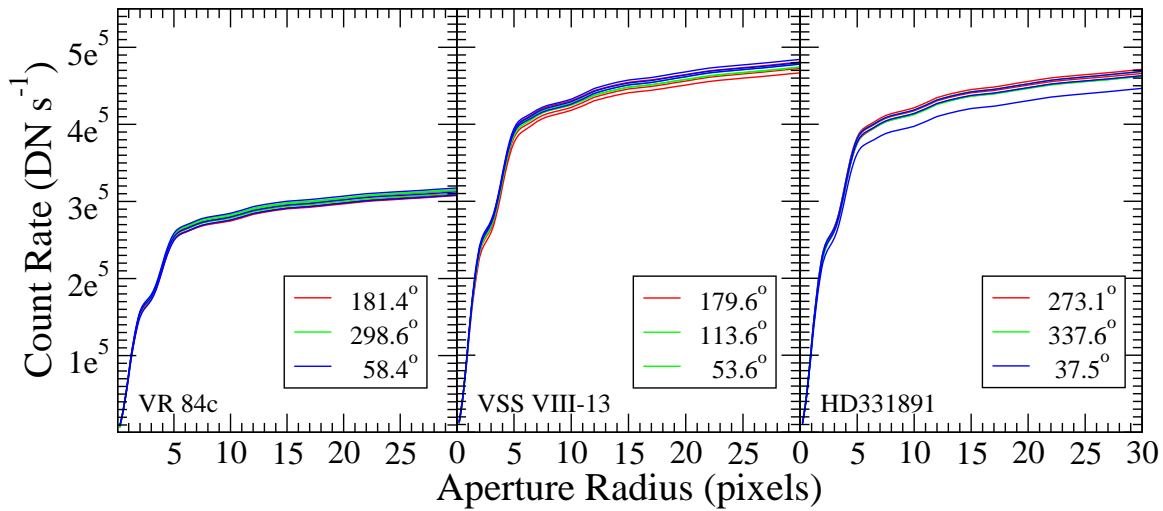}
\caption[Summed NIC2 radial profiles]{
For each target, the encircled count rate summed over all dither positions, for each polarizer and target field celestial orientation, is shown. 
The inflexions due to the PSF are clearly visible, as are some deviant profiles (e.g., HD 331891 at 37.5\degr). In all cases 95\% of the 
total flux is with a radius of 20 pixels, 99.9\% is at 50 pixels.
}\label{fig:rad2}
\end{figure*}

Comparison of NICMOS and ground-based polarization measurements must be band-transformed through the known wavelength dependence of interstellar polarization 
\citep{1975ApJ...196..261S}. The ground-based results are derived from observations at $1.21\micron$ and $2.04\micron$ and are corrected to $1.05\micron$ 
and $2.0\micron$ using the ``Serkowski curve'' \citep{1973IAUS...52..145S,1982AJ.....87..695W}. These values are listed in Table~\ref{tab:stand} and 
are used in this study as the ground-based comparison values. We also propagate the uncertainties of each ground-based measurement in our analysis.

\section{Data Reduction}\label{red}

The ``raw'' multi-accum exposures were used to create individual instrumentally calibrated count rate images for each target pointing. Calibration steps 
included: ``DC bias'' offset corrections with respect to the contemporaneously obtained reference dark frames, ``super-zero'' signal accumulation compensation, 
read-by-read dark subtraction, read-by-read linearity correction, comic-ray identification and reparation, saturation threshold determination,  flat-field 
correction,weighted-least squares count rate solution, bad pixel reparation by 2D interpolation of good neighbors and the removal of the telescope (+sky) thermal 
background. Before continuing, each image was checked for possible latent image persistence resulting from the previous image. No image latency was found in 
any of the individual multi-accum exposures. 

\subsection{Photometric Analysis}\label{phot}

It is essential to individually evaluate the fidelity of each pointed observation. An ideal imaging detector would produce precisely the same photometry with 
the target located at any arbitrary position on the grid of detector pixels, and would make polarimetric analysis fairly straightforward. Polarimetry, by its 
very nature, is highly sensitive to small variations in the photometry, and in any instrumental system there are many factors that can compromise the data. 
Sources of small variations in the photometry, that do not directly result from the intrinsic polarization of the target, and that have not already been 
accounted for, must be identified and removed. 

Aperture photometry was performed at each pointing using the {\tt digiphot} package within IRAF\footnote{IRAF is distributed by the National Optical Astronomy 
Observatories, which are operated by the Association of Universities for Research in Astronomy, Inc., under cooperative agreement with the National Science 
Foundation.}. Radially incremented circular apertures were placed on the target. The (sub-pixel) centers were determined using the intensity weighted means 
of the profiles in the x and y directions. The encircled count rates through the radially increasing apertures were then extracted. 

The summed extracted NIC2 radial profiles from the same polarizer at the separate position angles are presented in Figure~\ref{fig:rad2}. From this a number of 
effects are observed. Firstly, the Airy pattern component of the PSF is seen in the radial profile inflections. Secondly, there are some profiles that are 
inconsistent with the majority due to non-repeatability affects in the photometry beyond those mitigated by our observing strategy and image calibration. Finally, 
there is a noticeable spread in the count rates; more than one expects from polarization alone. 

\begin{figure}
\plotone{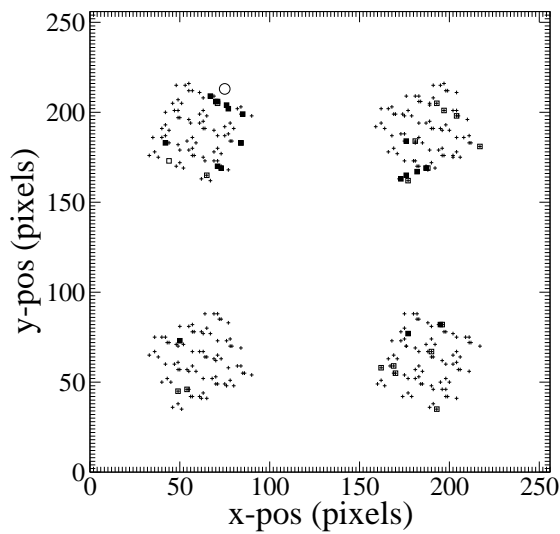}
\caption[Bad profiles]{
Identifying NIC2 deviant profiles. The open circle (r=0\farcs3, 4 pixels) marks the position and geometrical size of the coronagraphic hole projected onto 
the detector. However, the afocal imprint of the hole can affect 2.0\micron~circum-hole photometry (at our 2$\sigma$ rejection level) to a distance of 
$\sim0\farcs6$. ``+'' signs show the positions of all the pointings from all 9 orbits. The filled squares are profiles that show $>2\sigma$ deviations at 
all radii. The open squares are profiles that only show $>2\sigma$ deviations within a radius of two pixels. 
}\label{fig:badpoints}
\end{figure}

Each individual radial profile was compared to the average profile from the 12 individual dithers. The offset of each individual radial profile 
from the average profile was determined as a function of radius. Any encircled radial profile that showed more than a $2\sigma$ deviation from the average profile 
was flagged and inspected. Figure~\ref{fig:badpoints} spatially plots the positions of the flagged profiles using the filled squares. In all, 10\% of 
the profiles show deviations of $>2\sigma$ from the original average profile. 

About 40\% of the deviant profiles (4\% of all the pointings) are the result of the individual target pointings placing the center of the stellar 
PSF on uncorrectable bad pixels (excessive grot, i.e., pixels with reduced throughput due to particulate contamination on the detector\footnote{
http://www.stsci.edu/hst/nicmos/documents/isrs/isr\_99\_008.pdf}). In this case it is impossible to recover the lost flux and these pointings are 
discarded from further study. These profiles are typically clustered around the Q1-A positions. 

As expected, in $\sim20\%$ of the deviant cases the pointings have fallen close to the coronagraphic hole. The profiles are clustered around the Q2-A positions. 
With time, the precise position of the hole wanders by about one pixel. However, its position (x,y) is typically 73.5, 213.5 with a soft edge that can extend to a 
radius of 7 pixels. Data that were affected by the presence of the coronagraph were excluded. 

The remaining flagged profiles show $>2\sigma$ deviations at radii of $\leq2$ pixels. As the NIC2 polarizers have an effective wavelength of $1.995\micron$, and 
the effective primary aperture is 2.281~m (as defined by the NIC2 pupil mask), the theoretical PSF core diameter is 0\farcs22 or 2.91 pixels (PSF FWHM = 0\farcs185, 
2.44 pixels); NIC2 critically samples the PSF. As these profiles are only deviant within the diffraction limit of the instrument, and are within $2\sigma$ of 
the average profiles at radii $>2$ pixels, i.e., the effects are negated by increasing the aperture size, they are not excluded from further study. 

The processes of flagging and inspecting deviant profiles was repeated until all profiles fell within $2\sigma$ of the average. At each iteration, excluded profiles 
were removed from the original average. In the worst case, only three profiles (from 12) were rejected. 

\subsection{Achievable Accuracies}

In polarimetric analyses, the accuracies that are achievable are determined by the signal-to-noise ratio (S/N) of the data. Therefore, in order to estimate the 
effects of background noise, the twelve NIC2 images in each polarizer were combined, after precise astrometric registration, into a single median image. In all 
cases, images were first aligned using the spacecraft pointing information provided in the {\tt .fits} headers of the raw files. This co-registered the images 
with a relative precision of a few tenths of a pixel. Subsequent ``fine'' alignments of the inter-visit images were performed using an apodized bi-cubic re-sampling 
of the stellar PSF cores onto a 32$\times$ (in both x and y) larger grid. The re-sampled individual images were then shifted to a common fiducial position by the 
difference in centroids found iteratively with least-squares 2D Gaussian profile fitting. This process was repeated three times and converged all re-registered 
image centroids to a few thousandths of a pixel.

\begin{figure}
\plotone{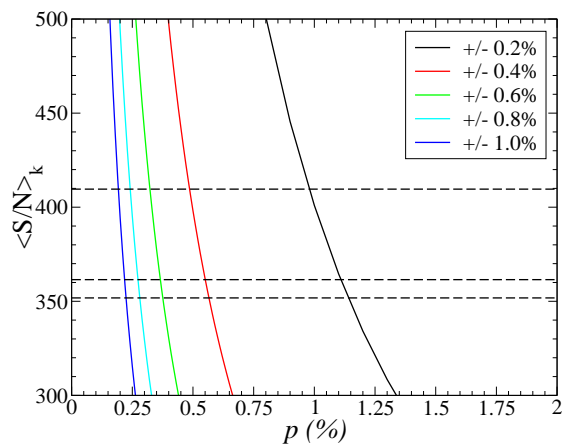}
\caption[Peak Pixel S/N for NIC2]{
Achievable polarimetric accuracies as a function of $\langle{\rm S/N}\rangle_k$ and intrinsic target polarization ($p(\%)$). The solid colored curves show the 
increasing uncertainties in $p$, given by the inset, as $p$ approaches zero for a given $\langle{\rm S/N}\rangle_k$. The dashed horizontal lines, from top to 
bottom, are the peak per pixel $\langle{\rm S/N}\rangle_k$ for VR 84c, VSS VIII-13, and HD331891 respectively, averaged across all three polarizers. For example, 
using the data collected on VR84c, we expect to reach accuracies of $p=1.0\pm0.2\%$ or $0.5\pm0.4\%$ depending on the actual intrinsic polarization of VR84c. 
Polarizations below $0.35\%$ will be consistent with zero as the uncertainties are $\pm0.6\%$. 
}\label{fig:peaksn}
\end{figure}

The pixel-to-pixel noise in the combined polarimetric total intensity images (Stokes $I$) was found by measuring (in each image) the standard deviation in the 
``background'' region. The region was close enough to the target that all 12 backgrounds from the individual images contributed to the combined image, but 
sufficiently far out so no flux from the target itself contributed to the background. A $151\times151$ pixel sub-array region, centered on the target was used. 
Flux from the diffraction spikes was masked. To find the region beyond which the target did not contribute to the background, square masks, increasing in size 
iteratively, were placed in the photometric aperture. As the mask was incremented (in 2 pixel lengths and widths) from 90 to 100 pixels, earlier (smaller 
mask size) standard deviations ceased to decrease, and both the median and mean were consistent with zero. Therefore, a square mask of $100\times100$ pixels, 
plus a diffraction spike mask, were used and noise statistics were computed on 10660 pixels. In all cases, the peak pixels (with instrument intensities of 
several thousand counts per second) typically have a $1\sigma$ noise of 0.1 to 0.2 counts per second; the per pixel data is photon noise dominated. 

As shown by \citet[][hereafter SA99]{1999PASP..111.1298S}, the S/N ratio averaged across the three polarizers ($\langle{\rm S/N}\rangle_k$) times the required 
polarization degree ($p$) determines the accuracy to which polarimetry can be performed ($\sigma_p$) in the photon noise dominated regime. Equation~\ref{equ:sanda}, 
taken from SA99 (their section 7.4), demonstrates the relationship between $\langle{\rm S/N}\rangle_k$ and $\sigma_p$. It does not take into account any instrumental 
polarization, nor any differences in the PSFs between polarizers, and assumes that the polarizers are perfect. 

\begin{equation}\label{equ:sanda}
\log_{10}(\sigma_p/p) = -0.102 -0.9898\log_{10}(p\langle{\rm S/N}\rangle_k)
\end{equation} 

Given $p$ and S/N we can use Equation~\ref{equ:sanda} to calculate the theoretical achievable polarimetric accuracies in the absence of non photon noise dominated 
statistics, as illustrated in Figure~\ref{fig:peaksn}. It can be seen (by looking at the intercepts of the highest dashed horizontal line and the $\pm0.2\%$ curve) 
that the observations theoretically allow measurements of $p=1.0\pm0.2\%$ for VR84c. 

However, as a single NIC2 pixel spatially under-samples the  PSF through the polarizers by 59\%, accurate polarimetry cannot be done on a single pixel; the image 
plane of NIC2 better than critically (Nyquist) samples the FWHM of the PSF. In addition, there are temporal instabilities in the PSF from spacecraft breathing 
and pupil mask shifts (the three polarizers are used non simultaneously). For point-source polarimetry, some spatial binning beyond an instrumental resolution 
element must be done to improve the accuracy of the polarimetry otherwise affected by the aforementioned instabilities and measurement limitations. 

So far we have assumed that the intrinsic pixel calibration uncertainties are largely mitigated through an effective image dithering strategy, as was implemented 
for NIC2 in this calibration program. However, there remains a clear dispersion in the photometry after dither combination. As expected, this dispersion varies with 
radius because the diffracted (and scattered) energy extends very far from the central pixel (Figure~\ref{fig:rad2}). In addition, for any fixed radius, the dispersion 
is also separately affected by target position on the array due to instrumental affects (e.g., imperfect flat fielding and IPRFs that, in detail, differ among detector
photodiodes/pixels). 

\begin{figure}
\plotone{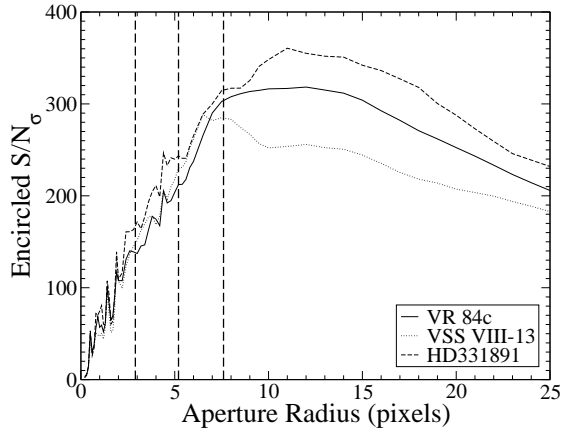}
\caption[Photometric Dispersion S/N]{
The radial variation of ${\rm S/N_\sigma}$ in each target. ${\rm S/N_\sigma}$ is defined by the values of the average profiles divided by the $1\sigma$ dispersions derived 
from the individual profiles (see text for more details). The three thick vertical dashed lines show the theoretical radii of the $1^{\rm st}, 2^{\rm nd}$ and 
$3^{\rm rd}$ Airy minima arising from the diffraction pattern of the centrally obscured primary mirror.
}\label{fig:deviants}
\end{figure}

In the point source case we use the encircled energy curves, from the incrementing radius aperture photometry, to follow the variation in the S/N (and therefore 
the achievable accuracy). Small temporal variations in the PSF structure of point sources will be largely mitigated by polarimetric analyses done with measurements on 
sufficiently large target ``enclosing'' photometric apertures. The optimal size of the photometric measurement aperture will be large enough to minimize 
the aggregate affects of changes at the pixel level, but not so large as to include data in the noise regime that is not photon dominated. Therefore, the S/N from 
dithered images is defined as the ratio between the mean of the individual sums of per-pixel measures in each image, and the $1\sigma$ (standard deviation) of the 
sums about that mean (${\rm S/N_\sigma}$). With this definition we can see the polarimetric accuracy possible considering instrumental and telescope systematic effects 
(quadrant dependence, IPRF, PSF, telescope breathing, etc).

Figure~\ref{fig:deviants} shows the NIC2 ${\rm S/N_\sigma}$ as determined from the $1\sigma$ dispersions about the mean in the incremental-radius aperture photometry. 
The typical ${\rm S/N_\sigma}$ that encircles the $2^{\rm nd}$ Airy maxima is 300. From Figure~\ref{fig:peaksn} we see that a polarization measurement of $p=1.3\pm0.2\%$ is 
achievable. This is consistent with the pixel-to-pixel $\langle{\rm S/N}\rangle_k$ accuracy. The ${\rm S/N_\sigma}$ that encircles the $1^{\rm st}$  Airy maxima 
gives a possible polarization measure of $p\approx1.2\pm0.3\%$, inside the first Airy minimum we see $p\approx1.1\pm0.5\%$ is attainable. Apertures with radii of 
7.6 pixels are optimum. Inside this radius ${\rm S/N_\sigma}$ is dominated by changes at the pixel level (large variations with radius), outside this radius photon noise 
begins to dominate (${\rm S/N_\sigma}$ smoothes out and turns over). At even larger radii, the target flux will strongly decline so that instrumental noise will dominate. 

The achievable accuracies are optimized by using a photometric aperture size of 7.6 pixels (0\farcs58). Moreover, in \S~\ref{calib} we will demonstrate that by re-calibration 
of the polarizer transmission coefficients, using our new data set, we can bring the observed values of $p$ and the celestial orientation of polarization ($\theta$) in 
agreement with the ground-based measurement of the polarized standard stars. However, as demonstrated by Figure~\ref{fig:peaksn}, in the case of the un-polarized standard, 
we cannot expect to determine $p=0\%$, but merely a polarization that is consistent with zero from within the errors. 

\section{Polarimetric Analysis}\label{ptheta}

The determination of $p$ and $\theta$ have been specifically addressed for NIC2 previously \citep[B06;][]{2000PASP..112..983H}. 
The coefficients presented in Table~\ref{tab:coeffs} (the best available previous determination without the addition of our newly 
acquired calibration data) are those determined from the Cycle 11 NICMOS polarimetric calibration program \citep{2002hstc.conf..258H}.  
We briefly revisit the linear technique for the case of three non-ideal polarizers below; the case of three ``ideal'' polarizers 
(applicable for ACS, but not for NICMOS) is discussed by \cite{2007A&A...471..137C}.

\begin{deluxetable}{lcccc}
\tablecaption{The Previous Polarimetric coefficients\label{tab:coeffs}}
\tablewidth{0pt}
\tablehead{
\colhead{Polarizer}&\colhead{$\phi_k$ (\degr)}&\colhead{$\eta_k$}&\colhead{$l_k$}&\colhead{$t_k$}
}
\startdata
POL0L   & 8.84   & 0.7313 & 0.1552 & 0.8779 \\
POL120L & 131.42 & 0.6288 & 0.2279 & 0.8379 \\
POL240L & 248.18 & 0.8738 & 0.0673 & 0.9667 \\
\enddata
\tablecomments{
The previous polarization coefficients as derived from the Cycle 11 calibration program. 
}
\end{deluxetable}

\begin{figure*}
\plotone{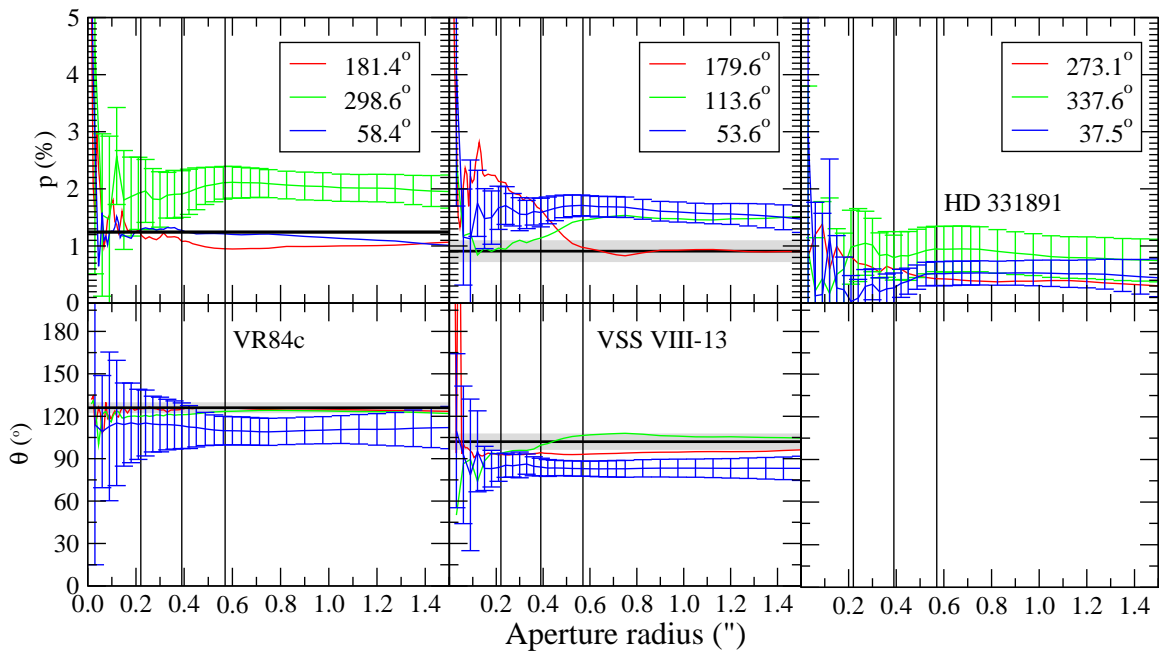}
\caption[Previous Calibration]{
The results using the previous calibration. The thick solid lines show the ground based results and the grey shaded areas show the uncertainties 
associated with those ground-based results.  To avoid cluttering the plot, error bars are only shown for the cases where they do not overlap the 
ground based results. The vertical solid lines mark the radii of the $1^{\rm st}, 2^{\rm nd}$ and $3^{\rm rd}$ dark minima of the PSF. The values 
of $\theta$ are in the celestial reference frame (position angle east of north).  The values of $\theta$ for HD 331891 bear no meaning and have 
therefore been excluded. 
}\label{fig:oldcal}
\end{figure*}

\begin{deluxetable*}{lcccccc}
\tablecaption{The NIC2 Calibration from Multiple Orientations\label{tab:rollsnic2}}
\tablecolumns{7}
\tablewidth{0pt}
\tablehead{
\colhead{Target}&\colhead{Polarizer}&\colhead{$I_1$}&\colhead{$I_2$}&\colhead{$I_3$}&\colhead{$p\%$}&\colhead{$\theta(^0)$}             
}
\startdata
VR 84c      & POL0L   & 22357.5 & 22855.3 & 22578.1 & $1.7\pm0.3$ & $115\pm2$ \\
            & POL120L & 22914.5 & 22751.2 & 22721.1 & $0.8\pm0.3$ & $126\pm6$ \\
            & POL240L & 22775.4	& 22650.3 & 23156.0 & $1.6\pm0.5$ & $122\pm6$ \\
VSS VIII-13 & POL0L   & 34362.8 & 34586.6 & 34804.0 & $1.1\pm0.3$ & $79\pm10$ \\
            & POL120L & 34733.7 & 34816.8 & 34635.1 & $0.5\pm0.2$ & $85\pm22$ \\
            & POL240L & 34969.2 & 34487.8 & 35190.2 & $1.3\pm0.3$ & $105\pm6$ \\
HD 331891   & POL0L   & 33817.0 & 33881.5 & 33636.2 & $0.6\pm0.6$ & \nodata   \\
            & POL120L & 34154.7 & 34081.5 & 33929.6 & $0.7\pm0.2$ & \nodata   \\
            & POL240L & 34244.6 & 34063.8 & 34054.5 & $0.4\pm0.2$ & \nodata   \\
\enddata
\tablecomments{
All results have been extracted from the apertures with a radius of 7.6 pixels (0\farcs58) and combines the data from the individual polarizers at the 
separate roll angles. $I_1, I_2$ and $I_3$ are in units of counts per second, averaged across the dithers, per polarizer, per orbit. 
}
\end{deluxetable*}

The instrumental counts per second measured through each polarizing element are used to define an observed intensity vector of the form 
$a=[I_1,I_2,I_3]$. The Stokes parameters, which  also defined a vector $b=[I,Q,U]$, are used to calculate $p$ and $\theta_f$ (see Eqs. 
\ref{equ:p} and \ref{equ:theta}). The two vectors $a$ and $b$ are simply related to each other by the linear expression $[C]b=a$, where 
$[C]$ is a matrix describing the characteristics of the $k^{\rm th}$ polarizer, namely the actual orientation (in radians) of the polarizer 
($\phi_k$), the fraction of the light transmitted in the parallel direction ($t_k$), and the fraction of light transmitted in the perpendicular 
direction (the ``leak'', $l_k$). The polarizer efficiency ($\eta_k$) is given by $(1-l_k)/(1+l_k)$. The linear expression $[C]b=a$ 
can be solved for $(I,Q,U)$ using LU decomposition. We can then determine $p$ and $\theta_f$ using Equations~\ref{equ:p} and \ref{equ:theta}.

\begin{equation}\label{equ:p}
p = 100\% \times \frac{\sqrt{Q^2+U^2}}{I}
\end{equation}

\begin{equation}\label{equ:theta}
\theta_f = \frac{1}{2}\arctan{\frac{U}{Q}}
\end{equation}

In Equation~\ref{equ:theta} a 360\degr~arctangent function is assumed. In addition, the orientation of the frame has to be subtracted from $\theta_f$ in order to retrieve 
the celestial position angle $\theta$. This process has been coded into the IDL routine {\it polarize.pro}\footnote{http://www.stsci.edu/hst/nicmos/tools/polarize\_tools.html}. 
This routine takes the three polarized images, calculates $(I,Q,U)$, and then produces two dimensional maps of $p$ and $\theta_f$. The $1\sigma$ dispersions of the 
count rates around the mean (used to define ${\rm S/N_\sigma}$) have also been used to determine the uncertainties in $p$ and $\theta$. As shown by SA99, the variance and 
covariance in the $(Q,U)$ plane define the uncertainties in $p$ ($\sigma(p)$) and $\theta$ ($\sigma(\theta)$). 

\begin{figure*}
\plotone{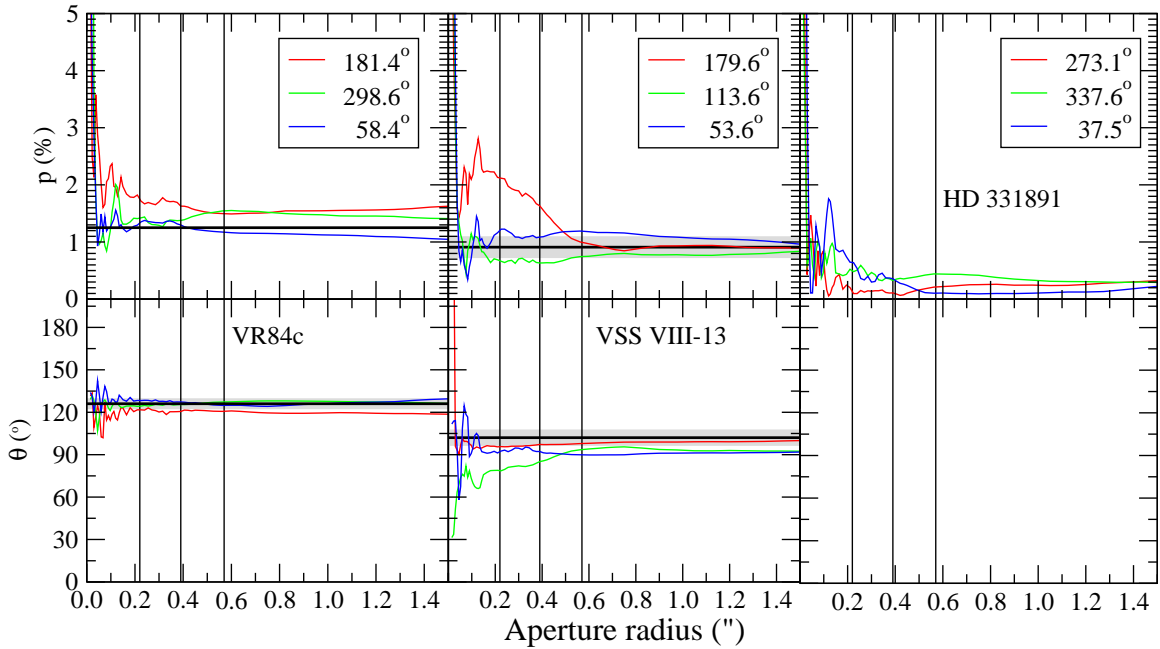}
\caption[The New Calibration]{
The results using the new calibration. The same convention is used as for Figure~\ref{fig:oldcal}. All results are now consistent with the ground-based data. The values of 
$\theta$ for HD 331891 bear no meaning and have therefore been excluded. 
}\label{fig:newcal}
\end{figure*}

\begin{deluxetable*}{lcccccccc}
\tablecaption{The NIC2 Instrumental\label{tab:innic2}}
\tablecolumns{7}
\tablewidth{0pt}
\tablehead{
\colhead{Target}&\colhead{Polarizer}&\colhead{$I_{nic}$}&\colhead{$Q_{nic}$}&\colhead{$U_{nic}$}&\colhead{$Q_{std}$}&\colhead{$U_{std}$}&\colhead{$p_{ins}\%$}&\colhead{$\theta_{ins}(^0)$}
}
\startdata
VR 84c      & POL0L   & 44574.1 &  -490.7 & -573.8 & -172.2 & -529.9 & $0.7$ & $94$ \\
            & POL120L & 44315.3 &  -112.7 & -355.4 & -171.2 & -526.8 & $0.4$ & $126$ \\
            & POL240L & 44294.8 &  -309.4 & -613.7 & -171.1 & -526.6 & $0.4$ & $106$ \\
VSS VIII-13 & POL0L   & 68183.2 &  -680.6 &  281.6 & -566.8 & -252.4 & $0.8$ & $141$ \\
            & POL120L & 67523.4 &  -325.9 &   52.6 & -561.3 & -249.9 & $0.6$ & $116$ \\
            & POL240L & 67563.3 &  -787.7 & -439.9 & -561.7 & -250.1 & $0.4$ & $110$ \\
HD 331891   & POL0L   & 66628.9 &    85.1 & -399.8 &    0.0 &    0.0 & $0.6$ & \nodata   \\
            & POL120L & 66209.9 &  -315.8 &  297.2 &    0.0 &    0.0 & $0.7$ & \nodata   \\
            & POL240L & 66142.9 &   249.9 & -121.8 &    0.0 &    0.0 & $0.4$ & \nodata   \\
\enddata
\tablecomments{
The results from the NIC2 instrumental polarization tests. All results have been extracted from the apertures with a radius of 7.6 pixels. 
}
\end{deluxetable*}

In order to demonstrate the limiting accuracy of the previous (Cycle 11) polarimetric calibration, we have applied the Table~\ref{tab:coeffs} coefficients to our newly acquired 
polarimetric standard star observations. Figure~\ref{fig:oldcal} demonstrates that only at a few roll angles does this reproduce the ground-based results. In all cases, however, 
we can see that the results are stable outside of the $2^{\rm nd}$ Airy maximum.  

\subsection{The Calibration}\label{calib}

It is now that we can begin to exploit the power of observing each standard at multiple spacecraft roll angles; the uncertainties due to the relative filter transmissions 
can be removed as the same polarizers have been used at three distinct orientations, i.e., each polarizer has separately, and independently, collected all of the necessary 
polarimetric data. By using polarimetric data gathered in this fashion, the method for calculating $p$ and $\theta$ can use invariant transmission coefficients for a 
given polarizer. In this case the resulting polarization is independent of $t_k$ and is only affected by $\phi_k$ (plus the ORIENTAT) and $l_k$. Due to observation scheduling 
constraints imposed by {\it HST}'s 2-Gyro operating mode (in place during Cycle 15) we were unable to obtain our calibration data at ideal differential celestial position angles 
(see Table~\ref{tab:obs} for as-executed orientations). Hence, though obtained with three roll angles, we cannot analyze these data as one would through a single 
perfect polarizer at ideal orientations; the method described in \S~\ref{ptheta} must still be followed in order to derive the Stokes parameters. 

The quantitative results obtained from polarimetric analysis of the apertures enclosing the $2^{\rm nd}$ Airy maxima, from the three roll orientations using the Cycle 11 
derived calibration coefficients (Table~\ref{tab:coeffs}), are shown in Table~\ref{tab:rollsnic2}. These results do not precisely recover the ground-based results listed in 
Table~\ref{tab:stand}. With the unknowns ($t_k$) nullified by the three orientations, we assume these residuals to be an instrumental polarization ($p_{ins}$). 

Since $(Q,U)_{ins}$ is in the instrumental frame, the derived $p_{ins}$ is fixed to the co-rotating spacecraft frame. Therefore, simply subtracting the observed residual 
polarizations (found in the unpolarized standard) from the polarized standards is not appropriate; $p_{ins}$ must be determined in Stokes space. For this, as we are 
calibrating NICMOS against the ground-based results, we determine $(I,Q,U)$ (and the associated uncertainties) for the standard stars, $(I,Q,U)_{std}$. However, $I$ is 
dependent on the nature of each individual observation and the intrinsic polarization of the target, so we must calculate $(Q,U)_{std}$ as a function of the $I$ observed 
by NIC2 ($I_{nic}$). It then follows that we can determine $(Q,U)_{ins}$ from Equation~\ref{equ:qui} (where all vectors have been rotated into the celestial plane). 
We can then use Equations~\ref{equ:p} and ~\ref{equ:theta} to determine $(p,\theta)_{ins}$. The results of this are presented in Table~\ref{tab:innic2}, which shows 
the raw Stokes parameters derived from following this method. The instrumental polarization derived from an unweighted average of our observations is $0.6\pm0.1\%$ at 
$116\pm15\degr$. 

\begin{equation}\label{equ:qui}
(Q,U)_{ins} = (Q,U)_{nic} - (Q,U)_{std}
\end{equation}

Attention can now be turned back to the data from the single roll angles. With the instrumental polarization (in Stokes space) determined (and subtracted), the only parameter 
affecting the deviation from the ground-based results is $t_k$. The values of $t_0$ and $t_{120}$ that reproduce the ground-based results can then be determined numerically. 
As in the previous calibrations, $t_{240}$ is held at 0.9667 as it is the long polarizer with the highest efficiency. The other values of $t_k$ can be determined with respects 
to $t_{240}$. 

As this process was performed by B06 on archival data, we can also include the values of $t_k$ derived by those authors. We limit these additional data to the polarimetric 
standards observed in the post-NCS era. These data comprise of two epochs for VR 84c (identified as CHA-DC-F7 by B06) and two epochs of data for HD 331891. Table~\ref{tab:newtks} 
presents the resulting values of $t_0$ and $t_{120}$ including the derived values of $p$ and $\theta$. The un-weighted average values of $t_0$ and $t_{120}$ are $0.882\pm0.008$ 
and $0.837\pm0.004$ respectively. 

\begin{deluxetable}{lccccccc}
\tablecaption{Adjusting the Transmission Coefficients.\label{tab:newtks}}
\tablecolumns{6}
\tablewidth{0pt}
\tablehead{
\colhead{Target}&\colhead{PA}&\colhead{$t_0$}&\colhead{$t_{120}$}&\colhead{p (\%)}&\colhead{$\theta$}\\
}
\startdata
VR84c       & -178.6\degr & 0.8827 & 0.8346 & $1.2\pm0.4$ & $126\pm3$ \\
            &  -61.4\degr & 0.8804 & 0.8369 & $1.3\pm0.4$ & $126\pm5$ \\
            &   58.4\degr & 0.8883 & 0.8447 & $1.3\pm0.4$ & $127\pm7$ \\
(B06)       &  -50.8\degr & 0.8914 & 0.8372 & $1.2\pm0.1$ & $125\pm7$ \\
(B06)       & -160.8\degr & 0.8897 & 0.8397 & $1.2\pm0.1$ & $124\pm6$ \\
VSS VIII-13 &  179.6\degr & 0.8737 & 0.8335 & $0.9\pm0.4$ & $103\pm12$\\
            &  113.6\degr & 0.8771 & 0.8369 & $0.9\pm0.4$ & $103\pm13$\\
            &   53.6\degr & 0.8916 & 0.8425 & $0.8\pm0.3$ & $102\pm8$ \\
HD 331891   &  -86.9\degr & 0.8804 & 0.8391 & $0.3\pm0.4$ & \nodata   \\
            &  -22.4\degr & 0.8860 & 0.8302 & $0.5\pm0.5$ & \nodata   \\
            &   37.5\degr & 0.8849 & 0.8391 & $0.4\pm0.4$ & \nodata   \\
(B06)	    & -109.6\degr & 0.8629 & 0.8269 & $0.3\pm0.1$ & \nodata   \\
(B06)       &    0.6\degr & 0.8718 & 0.8334 & $0.3\pm0.1$ & \nodata   \\  
\enddata
\tablecomments{
Determining the parallel transmission coefficients that reproduce the ground-based results (taken from a 7.6 pixel aperture). 
}
\end{deluxetable}

\begin{figure}
\plotone{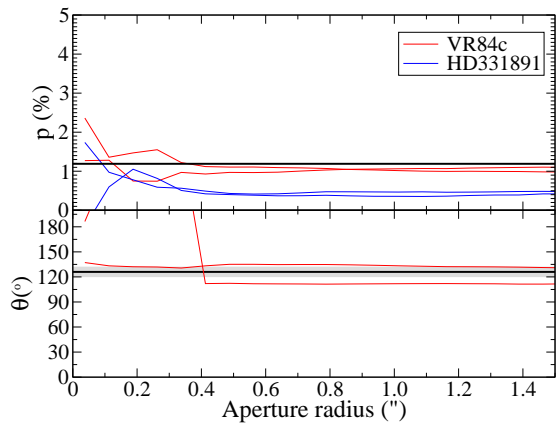}
\caption[Checking the new calibration against the old data]{
Checking the new calibration against the data that originally suggested an instrumental polarization. The errors bring all consistent with the ground-based results. 
The values of $\theta$ for HD331891 have been omitted for clarity, and because they bear no meaning. 
}\label{fig:oldcheck}
\end{figure}

We apply these newly re-derived coefficients to ascertain ($p$,$\theta$) for our target stars with the results shown in Figure~\ref{fig:newcal} (comparable directly to 
Figure~\ref{fig:oldcal}) and find the new calibration produces results consistent with the ground-based determinations in all cases. As an additional check we also apply 
the updated transmission coefficients to the data presented by B06 that suggested the original instrumental polarization. Figure~\ref{fig:oldcheck} plots the results from these 
re-analyzed data (it is similar to Figure 2 from B06). As can be seen, re-analysis with the now improved determinations of the polarizer transmissivities produce ($p,\theta$) 
results that are also now consistent with the ground-based data. The typical uncertainties associated with the profiles in Figure~\ref{fig:oldcheck} are $\sigma_p\approx1\%$ 
and $\sigma_\theta\approx10\degr$. The Cycle 11 data were not gathered using the additional dither points employed by this project; the polarimetric analysis performed on the 
archival data was improved through our updated calibration.  

\section{Discussions}\label{dis}
 
We have used our data to perform a detailed investigation of the NIC2 instrumental polarization. This work was motivated, in part, by recent attempts to observe 
objects that have intrinsic polarizations of less than 5\%. The previous calibrations were of sufficient accuracy to complete studies of more highly polarized targets 
but could not be used for high accuracy polarimetry of targets with low intrinsic polarization. By obtaining high S/N observations of one unpolarized and two polarized 
standard stars, at three well separated roll angles, we have been able to determine the level and orientation of the NIC2 instrumental polarization. By subtracting the 
instrumental polarization, in Stokes space, and improving upon the previous determination of the parallel transmission coefficients, we have successfully reproduced the 
polarimetric parameters of three standard stars observed from the ground. These results were consistent for each standard observed at three different instrumental 
orientations in the celestial frame. Our results make this the most comprehensive polarimetric calibration of NIC2 to date (and quite likely the last) for direct imaging. 
We suggest this quantitative characterization of the NIC2 polarimetric system supersede the previous post-NCS era polarimetric calibrations.

As the techniques used here require extremely precise photometry, there are many potential sources contributing to the uncertainties in determining $p$ and $\theta$. 
Highly precise photometric repeatability is limited by temporal, thermal, and metrologic optical instabilities in the {\it HST}+NICMOS+ instrumental system affecting 
the PSF, as well as spatial quantization and performance defects in the NIC2 detector and readout electronics. The measurement dispersions introduced by these (and 
perhaps other systematic) effects are carried through the SA99 analysis to determine $\sigma(p)$ and $\sigma(\theta)$. However, as we have been able to determine 
revised values for $t_0$ and $t_{120}$ in 13 cases, we can assign uncertainties to these coefficients. By propagating these uncertainties through the polarimetric 
analysis we find that they dominate over the errors derived following SA99. For example, applying the new coefficients to VR84c we find $p=1.3\pm0.4, \theta=126\pm5\degr$ 
following SA99, but by propagating the uncertainties in $t_k$ we find $p=1.3(+0.7,-0.6)\%, \theta=126(+10,-24)\degr$. 

The remaining uncertainties in the derived transmission coefficients do indeed dominate the statistical uncertainties from the photometry. What affect do these 
uncertainties have on the observationally determined values of $p_{ins}$ and $\theta_{ins}$? Could these uncertainties be responsible for the observed instrumental polarization? 
As can be seen from Table~\ref{tab:rollsnic2}, deviations between $I_1, I_2$ and $I_3$ (that in an ideal detector would be invariant) are no greater than $\sim2\%$ which 
corresponds to a $1\sigma$ error in $p_{ins}$ of only 0.1\%. However, a more accurate way to determine $p_{ins}$ requires observations of an unpolarized standard star across the 
whole detector area to map any field dependence in the polarization. As our results have been derived from dithered observations, if there is a field dependence in the polarization, 
then that could mimic an instrumental polarization on our data. Such a study cannot be done with the current data, so it is not possible to assert $p_{ins}$ at such a level. Instead, 
it is more appropriate to present the 0.6\% value as an instrumental upper limit; $p_{ins}$ could indeed be zero. 

What then happens to the rest of our analysis if $p_{ins}$ is indeed zero? To test this, we have re-derived the values of $t_k$ without subtracting $p_{ins}$ in the ($Q,U$) 
plane. Taking this approach we find $t_0=0.884\pm0.008$ and $t_{120}=0.838\pm0.003$ which is entirely consistent with the results taken from the instrumentally subtracted $(Q,U)$. 
So, again we find that the errors in $t_k$ dominate the uncertainties in the measured values of $p$ and $\theta$. As these uncertainties have been derived from 13 separate 
observations, that are all subjected to the photometric uncertainties themselves, we are left with the understanding that this calibration is limited by the photometric 
repeatability of individual observations. This includes all of the known affects presented in \S~\ref{obs}.

When comparing the new parallel transmission coefficients with the old it can be seen that $t_{120}$ is consistent with both calibrations; only $t_0$ has needed to be 
changed, albeit ``only'' by $\sim0.5\%$. Applying this correction to previous studies of {\it highly} polarized targets will therefore not alter the results by any significant amounts. 
However, when applied to targets with low polarization this ``tweak'' has a tremendous affect.

\subsection{Recommended Observing Strategy}\label{dis2}

In the cases where the target is spatially resolved, and does not contain a strong point source, then observers should bin their data to $3\times3$ pixels (in order to 
meet the diffraction limit of the instrument). Dithering can be used in the usual way to increase spatial resolution as well as to mitigate bad pixel affects. However, 
signal to noise will still limit the accuracy of $p$ and $\theta$ determinations, especially in targets whose surface brightnesses vary significantly on small spatial 
scales, i.e., comparable to, or smaller than, the diffraction limit. 

The case of a spatially unresolved source is not so straightforward. As explained above, the uncertainties in $t_k$ do dominate the accuracy of $p$ and $\theta$. However, as this 
study has shown, non-repeatabilities in aperture photometry (in the absence of well-dithered observations) suggest that unidentified outliers can readily bias polarimetric analysis 
independent of the intrinsic accuracy of the  polarimetric calibration. The data have been carefully inspected to identify photometrically deficient pixels by comparing each individual 
profile to the dispersion in the photometry measured from $\sim12$ pointings. This study has shown that $\sim10\%$ of all target pointings (even those avoiding known photometrically 
deficient areas of the detector like the coronagraphic hole) will include uncorrectable bad pixels that affect the measured polarization. A method to robustly detect, and reject, 
photometric measures degraded by such effects is to perform a target raster with a dither pattern with three or more pointings. The step size should be more than three times greater 
than the FWHM of the PSF to avoid persistence. Encircled energy (intensity) profiles should be derived independently from each observation (dither point). With a sufficiently large 
number of dither points, a sigma clip can then be used to determine whether an individual profile is deviant or not. The greater number of dithers used, the more accurate this clip 
will be. For small numbers of dither points (not optimally recommended) medianing can be used to coarsely reject outliers. The average of the remaining profiles can then be used to 
derive $p$ and $\theta$ from the coefficients presented in Table~\ref{tab:newcoeffs}. 

\begin{deluxetable}{lcccc}
\tablecaption{The New Polarimetric coefficients\label{tab:newcoeffs}}
\tablewidth{0pt}
\tablehead{
\colhead{Polarizer}&\colhead{$\phi_k$ (\degr)}&\colhead{$\eta_k$}&\colhead{$l_k$}&\colhead{$t_k$}
}
\startdata
POL0L   & 8.84   & 0.7313 & 0.1552 & $0.882\pm0.008$ \\
POL120L & 131.42 & 0.6288 & 0.2279 & $0.837\pm0.004$ \\
POL240L & 248.18 & 0.8738 & 0.0673 & 0.9667 \\
\enddata
\tablecomments{
The new NIC2 polarization coefficients as derived by this program. 
}
\end{deluxetable}

This study has shown that the affects of the IPRFs are largely mitigated by effective dithering. However, the uncertainty in the calibration of $t_k$ ultimately dominates the accuracy 
of ($p,\theta$) polarimetric analysis for sources with intrinsically low polarization, if observed with a sufficient number of optimally placed dither positions; observers need not be 
concerned by the IPRF. Since observations through the three polarizers are not simultaneously obtained, temporal instabilities in the {\it HST}+NICMOS PSF are still a concern and the 
effects are readily seen in Figures~\ref{fig:rad2}, \ref{fig:deviants}, \ref{fig:oldcal} and \ref{fig:newcal}. Outside of an aperture of radius 0\farcs58, the PSF affects are seen 
to be alleviated. Inside this radius, the errors in $p$ and $\theta$ rise rapidly and it will be left to the discretion of the observer to weigh their required accuracy to the possible 
results of beam-depolarization. 

\section{Conclusions}\label{cons}

In a non-ideal imaging polarimeter, such as NICMOS, it is essential to observe several polarimetric standards, at three well separated position angles through each polarizer, 
in order to full characterize the instrumental polarization. The additional roll angles remove the uncertainties in the (unknown) parallel transmission coefficients and allow 
rotation of any instrumental polarization with respects to the equatorial frame. Using this technique we have placed an upper limit to the NIC2 instrumental polarization of 0.6\%. 
With a known value for the Stokes $I$ parameter, the instrumental polarization can be transformed into the $Q,U$ plane and be subtracted. New parallel transmission coefficients can 
then be determined numerically by comparing $p$ and $\theta$ with that of the apriori well-determined calibration standards. Following this approach we have determined 
the $t_0$ and $t_{120}$ coefficients to be $0.883\pm0.004$ and $0.837\pm0.004$ respectively. As with the previous calibrations of the NIC2 polarimeter, we held $t_{240}$ constant 
at 0.9667. The $t_{120}$ coefficient is consistent with the previous calibration, but our knowledge of $t_0$ has been improved, resulting in a change in its previously determined 
value by $\sim0.5\%$. Such a small change in $t_0$ does not warrant the re-analysis of previous NIC2 imaging polarimetry data of {\it highly} polarized ($p>5\%$) targets, but is 
significant for targets with intrinsically low polarization fractions.  

As we use 13 determinations of $t_k$, we are able to assign $1\sigma$ uncertainties to the calibration coefficients. Propagating these uncertainties through the polarimetric analysis, 
we find that they dominate all other sources of error. Applying these adjusted values of $t_k$ to this (and the archived) calibration data, we find that NIC2 is now capable of 
confidently detecting polarizations at a level of $\approx1.0\%$. The uncertainties associated with such measurements are $\pm0.6\%$ and $\pm15\degr$ in $p$ and $\theta$, for sources 
at the $\approx1\%$ level of intrinsic polarization. This is the first time that such a level of accuracy has been achieved with the NICMOS polarization calibration. This improved 
calibration opens a new domain for observational investigations with {\it HST} by enabling very high precision polarimetry of intrinsically very low polarization sources. 

\acknowledgments

We are grateful to the referee for their thorough reading of this manuscript. Support for Proposal number HST-GO-10839.01-A was provided by NASA through a grant from the Space 
Telescope Science Institute, which is operated by the Association of Universities for Research in Astronomy, Incorporated, under NASA contract NAS5-26555.

\appendix

\section{The NIC1 Calibration}\label{app:a}

For the NIC1 observations, the individual pointings for each polarizer were not dithered around the 12 point pattern used for the NIC2 observations. 
However, the observations from each separate polarizer were dithered so as to avoid persistence. The pointings were also placed on an area of NIC1 where the 
quantum efficiency (QE) gradients are particularly shallow and where the QE is reasonably high. This area was also sufficiently far from any ``grotty'' 
pixels and sufficiently far from the edge of the detector.  

\begin{figure}
\plotone{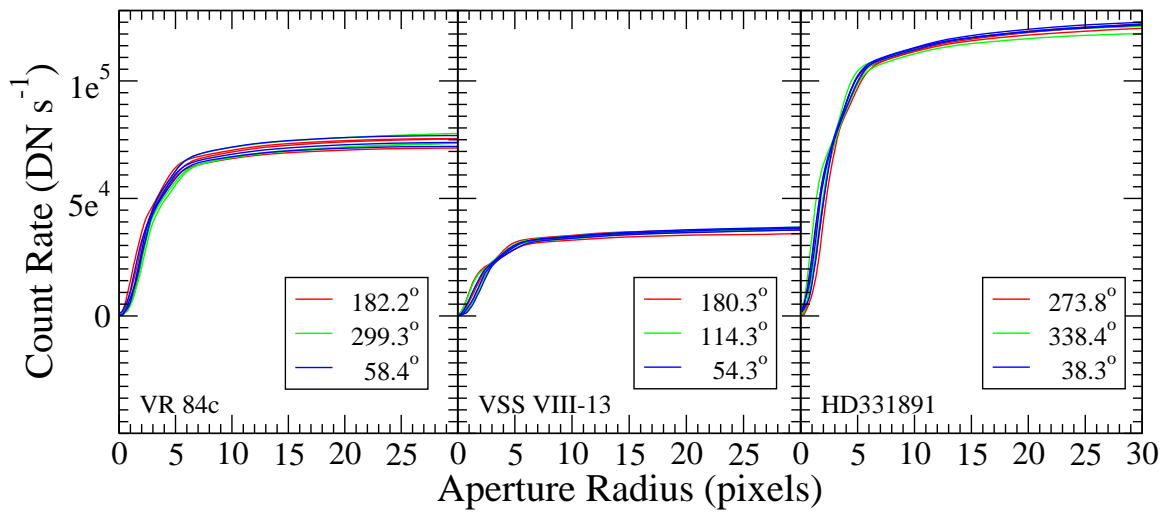}
\caption[NIC1 radial profiles]{
The NIC1 radial profiles. For each target, the encircled count rate for each polarizer and celestial orientation, is shown. 
}\label{fig:rad1}
\end{figure}

As there are only single pointings per polarizer, the data analysis for the NIC1 data is less complicated than for the NIC2 case. The extracted 
radial profiles from each polarizer (and pointing) at each roll angle are shown in Figure~\ref{fig:rad1}. For these profiles we can only define the S/N 
using the photon noise. As there are only single pointings in each NIC1 polarizer, it is impossible to perform the iterative $2\sigma$ clipping technique 
employed for NIC2. Therefore all NIC1 profiles must be included in the analysis. Although there is no statistical way to tell if the profiles in 
Figure~\ref{fig:rad1} are robust, we can at least see that all profiles are consistent with each other. This is likely due to the care taken in 
avoiding uncorrectable bad pixels. 

At the radius of the $3^{\rm rd}$ NIC1 dark Airy minimum (given by an aperture with $r=7.0$ pixels, assuming a central wavelength of $1.05\micron$ and 
a 2.281m aperture) the respective signal to noise ratios for VR84c, VSS VIII-13 and HD331891 are 257, 180 and 330 when averaged across the three polarizers. 
The theoretical NIC1 achievable accuracies are approximately $1.5\%$ with an uncertainty of $\pm0.2\%$ and $1.0\pm0.3\%$.

At 1.21\micron, \citet{1992ApJ...386..562W} give VR84c a polarization of $3.19\pm0.05\%$ at $118\pm1\degr$ and VSS VIII-13 a polarization of $2.21\pm0.3\%$ at 
$96\pm1\degr$. As there is no data for HD331891 at this wavelength we assume that this target is unpolarized. The polarizations and uncertainties are carried 
through the Serkowski correction and are presented in Table~\ref{tab:stand}.

Except for the $2\sigma$ clipping, the NIC1 data was treated exactly like the NIC2 data. The calibration data from the multiple orientations are presented in 
Table~\ref{tab:rollsnic1}. The NIC1 instrumental polarization was then determined to be $1.3\pm0.9\%$ at an orientation of $151\pm18\degr$. This instrumental 
polarization was subtracted in $(Q,U)$ space and the parallel transmission coefficients were re-derived on an orbit by orbit basis. The coefficients, and uncertainties, 
that bring all the data consistent with the ground-based data are $t_{120} = 0.57\pm0.01$ and $t_{240} = 0.69\pm0.01$. All the coefficients to be used in NIC1 
polarimetric observations are shown in Table~\ref{tab:n1coes}. Applying these new calibration coefficients to the original data
brings all polarimetric standards consistent with the ground-based results. 

\begin{deluxetable}{lccccccc}
\tablecaption{The NIC1 Calibration from Multiple Orientations\label{tab:rollsnic1}}
\tablecolumns{7}
\tablewidth{0pt}
\tablehead{
\colhead{Target}&\colhead{Polarizer}&\colhead{$I_1$}&\colhead{$I_2$}&\colhead{$I_3$}&\colhead{$p\%$}&\colhead{$\theta(^0)$}             
}
\startdata
VR84c       & POL0S   & 64292.5  & 68734.3  & 65147.3  & $4.1\pm0.1$    & $116.5\pm0.5$  \\
            & POL120S & 67785.9  & 64160.6  & 66039.8  & $6.47\pm 0.06$ & $131.6\pm0.6$  \\
            & POL240S & 66771.2  & 64206.3  & 68756.5  & $5.1\pm0.2$    & $122.8\pm0.2$  \\ 
VSS VIII-13 & POL0S   & 31156.2  & 32770.7  & 32472.4  & $3.1\pm0.2$    & $91\pm1$       \\
            & POL120S & 32393.0  & 32412.3  & 31762.6  & $3.0\pm0.1$    & $83\pm2$       \\
            & POL240S & 33123.9  & 31812.0  & 32638.4  & $2.9\pm0.2$    & $94\pm2$       \\ 
HD 331891   & POL0S   & 107931   & 107057   & 109394   & $1.3\pm0.1$    & \nodata        \\
            & POL120S & 108864   & 109376   & 108867   & $0.63\pm0.04$  & \nodata        \\
            & POL240S & 108221   & 108756   & 109656   & $1.06\pm0.06$  & \nodata        \\
\enddata
\tablecomments{
All results have been extracted from the apertures with a radius of 7.0 pixels and combines the data from the individual polarizers at the 
separate roll angles. 
}
\end{deluxetable}

\begin{deluxetable}{lcccc}
\tablecaption{The New NIC1 Polarimetric coefficients.\label{tab:n1coes}}
\tablewidth{0pt}
\tablehead{
\colhead{Polarizer}&\colhead{$\phi_k$ (\degr)}&\colhead{$\eta_k$}&\colhead{$l_k$}&\colhead{$t_k$}
}
\startdata
POL0S   & 1.42   & 0.9717 & 0.0144 & 0.7760  \\
POL120S & 116.30 & 0.4771 & 0.3540 & $0.57\pm0.01$ \\
POL240S & 258.72 & 0.7682 & 0.1311 & $0.69\pm0.01$ \\
\enddata
\tablecomments{
The new polarization coefficients, as derived by this program, for NIC1 polarimetry. 
}
\end{deluxetable}

Compared with the previous NIC1 calibration coefficients ($t_{120} = 0.5934$ and $t_{240} = 0.7173$) the coefficients derived here are lower. In the previous 
instance the coefficients were derived from observations of an unpolarized standard, and not from data at three well separated roll angles. However, the 
data used in this calibration have not been dithered over the detector (as in the case of the NIC2 data); instrumental affects will still be present. The 
fact that the uncertainties in the newly derived coefficients for NIC1 are larger than for the NIC2 case, reflects this finding. Propagating the coefficient 
uncertainties through the polarimetric analysis results in polarization measures of 5(+6,-2)\% at 123(+24,-12)\degr for VR84C, 3(+6,-2)\% at 94(+85,-27)\degr 
for VSS VIII-13 and 3(+5,-3)\% for HD331891. We therefore recommend that NIC1 only be considered for observations of targets that are postulated to have an 
intrinsic polarization of greater than 4\% per resolution element. The uncertainties in polarization measures from such a target will, however, be large. 

\bibliographystyle{apj}
\bibliography{batcheldor}

\end{document}